\documentclass[a4paper,fleqn,usenatbib]{mnras}

\usepackage{newtxtext,newtxmath}

\usepackage[T1]{fontenc}
\usepackage{ae,aecompl}

\usepackage{graphicx}	
\usepackage{amssymb}	
\usepackage{natbib}
\usepackage{soul}
\bibpunct{(}{)}{;}{a}{}{,}
\usepackage{graphicx}
\usepackage{multirow}

\usepackage{graphicx}
\usepackage{txfonts}
\usepackage{natbib}
\usepackage{xspace}
\usepackage{dcolumn}
\usepackage{longtable}
\usepackage{lscape}
\usepackage{soul}        

\usepackage{rotating}
\usepackage{afterpage}

\newcommand{\MJup}{M$_{\mathrm{Jup}}$\xspace}

\newcommand{\MSun}{M$_{\odot}$\xspace}
\newcommand{\LSun}{L$_{\odot}$\xspace}

\newcommand{\co}{$^{12}$CO}

\newcommand{\as}{\hbox{$^{\prime\prime}$}\xspace}

\title[The ALMA Early Science of V2775 Ori]{The ALMA Early Science view of FUor/EXor objects. I. Through the looking-glass of V2775 Ori\thanks{Based on ALMA observations, program number 2013.1.00710.S.}}

\author[A. Zurlo et al.]{Alice Zurlo$^{1,2,3}$,\thanks{E-mail: azurlo@das.uchile.cl}
Lucas A. Cieza$^{2,3}$, Jonathan P. Williams$^{4}$, Hector Canovas$^{5,3}$, \newauthor Sebastian Perez$^{1,3}$, Antonio Hales$^{6}$, Koraljka Mu\v{z}i\'c$^{2}$, David A. Principe$^{2,3}$, \newauthor Dary Ru\'iz-Rodr\'iguez$^7$, John Tobin$^8$, Yichen Zhang$^1$, Zhaohuan Zhu$^{9}$, Simon Casassus$^{1,3}$, \newauthor Jose L. Prieto$^{2}$  \\
$^{1}$Universidad de Chile, Camino el Observatorio 1515, Santiago, Chile \\
$^{2}$N\'ucleo de Astronom\'ia, Facultad de Ingenier\'ia, Universidad Diego Portales, Av. Ejercito 441, Santiago, Chile\\
$^{3}$Millenium Nucleus ``Protoplanetary Disks in ALMA Early Science'', Chile\\
$^{4}$Institute for Astronomy, University of Hawaii at Manoa, Honolulu, HI, 96822, USA\\
$^{5}$Departamento de F\'isica Te\'orica, Universidad Aut\'onoma de Madrid, Cantoblanco, 28049, Madrid, Spain\\
$^{6}$Atacama Large Millimeter/Submillimeter Array, Joint ALMA Observatory, Alonso de C\'ordova 3107, Vitacura 763-0355, Santiago, Chile\\
$^{7}$Research School of Astronomy and Astrophysics, Australian National University, Canberra, ACT 2611, Australia\\
$^8$Veni Fellow, Leiden Observatory, Leiden University, P.O. Box 9513, 2300-RA Leiden, The Netherlands\\
$^9$Department of Astrophysical Sciences, Princeton University, Princeton, NJ 08544, USA}

\date{Accepted 2016 November 2. Received 2016 November 2; in original form 2016 July 10}

\pubyear{2016}

\begin{document}
\label{firstpage}
\pagerange{\pageref{firstpage}--\pageref{lastpage}}
\maketitle

\begin{abstract}
   As part of an ALMA survey to study the origin of episodic accretion in young eruptive variables, we have observed the circumstellar environment of the star V2775 Ori. This object is a very young, pre-main sequence object which displays a large amplitude outburst characteristic of the FUor class.  We present Cycle-2 band 6 observations of V2775 Ori with a continuum and CO (2-1) isotopologue resolution of 0.25\as (103 au). We report the detection of a marginally resolved circumstellar disc in the ALMA continuum with an integrated flux of $106 \pm 2$ mJy, characteristic radius of $\sim$ 30 au, inclination of $14.0^{+7.8}_{-14.5}$ deg, and is oriented nearly face-on with respect to the plane of the sky.

The \co~emission is separated into distinct blue and red-shifted regions that appear to be rings or shells of expanding material from quasi-episodic outbursts.  The system is oriented in such a way that the disc is seen through the outflow remnant of V2775 Ori, which has an axis along our line-of-sight.  The $^{13}$CO emission displays similar structure to that of the \co, while the C$^{18}$O line emission is very weak. We calculated the expansion velocities of the low- and medium-density material with respect to the disc to be of -2.85 km s$^{-1}$ (blue), 4.4 km s$^{-1}$ (red) and -1.35 and 1.15 km s$^{-1}$ (for blue and red) and we derived the mass, momentum and kinetic energy of the expanding gas.  The outflow has an hourglass shape where the cavities are not seen. We interpret the shapes that the gas traces as cavities excavated by an ancient outflow. We report a detection of line emission from the circumstellar disc and derive a lower limit of the gas mass of 3 \MJup.
\end{abstract}




\section{Introduction}

FU Orionis stars (FUors) are variable, pre-main sequence low-mass stars that {  display} brightness variations on a very short timescale {  (less than one year)} \citep{1966VA......8..109H}. Their brightness can increase by 3-6 {  magnitudes} at optical wavelengths during the course of a few months and remain bright for decades, a phenomenon called ``outburst''. According to the episodic accretion scenario \citep{1996ARA&A..34..207H, 2014prpl.conf..387A}, most young stars are expected to experience these extreme luminosity outbursts during their early evolution.

The mechanism producing these sudden mass-accretion events, and thus the origin of these outbursts, is still poorly understood and currently there are several theories proposed: disc fragmentation \citep{2005ApJ...633L.137V, 2012ApJ...746..110Z}, thermal instability \citep{1994ApJ...427..987B}, coupling of gravitational and magneto-rotational instability \citep{2001MNRAS.324..705A, 2009ApJ...701..620Z}, and tidal interaction between the disc and a companion \citep{2004MNRAS.353..841L,1992ApJ...401L..31B}. During their ourbursts the bolometric luminosity of the FUor type stars is 50--500 $L_{\odot}$, and the accretion rate is between $10^{-6}$ and $10^{-4}$ \MSun yr$^{-1}$ \citep{2014prpl.conf..387A}. Over the full duration of an outburst the star may accrete $\sim$ 0.01 \MSun of material, roughly the mass of a typical T Tauri disc \citep{2005ApJ...631.1134A}. During this phase FUors show F-G supergiant optical spectra but with broader lines than those of T Tauri stars and K-M supergiant near-infrared spectra, showing CO overtone absorption \citep{2009ApJ...701..620Z}.
The Fe I, Li I, Ca I, and CO lines are double peaked and show broadening, as expected for a rotating disc kinematic \citep{1996ARA&A..34..207H}. {  S}ingle dish millimeter observations {  indicate} that FUors look more similar to Class I protostars rather than Class II stars \citep{2001ApJS..134..115S}. However, interferometric data with enhanced angular resolution and sensitivity are required to demonstrate this \citep[e.g.][]{2015ApJ...812..134H}. {  The outflows of these objects are driven by repeated outbursts with peak mass loss rate of $10^{-5}$ \MSun yr$^{-1}$ \citep{1994ApJ...424..793E}; the dusty environment of FUors stars shows a core/envelope region structure, in contrast to HAEBE stars which show a core structure \citep{1998A&A...336..565H}; and to well characterise the structure of FUors it is necessary to use interferometric (opposed to single dish) observations are necessary to characterize the circumstellar structure and environment of FUor objects \citep{2011A&A...535A.125K}.}

The FUor type star V2775 Ori was first identified by \citet{2011A&A...526L...1C} and is located in the Orion molecular clouds, near the southern edge of the L1641 region, at a distance of 414$\pm$7 pc \citep{2007A&A...474..515M}. \citet{2011A&A...526L...1C} describe it as a young stellar object (YSO) with characteristics of {  both} FUors and EXors (from the prototype FU Ori and EX Lup, respectively). EXors are pre-main sequence stars that display shorter timescale outbursts than the FUors. The spectrum of V2775 Ori shows a featureless continuum, with strong CO bands and H$_{2}$O broad-band absorption (typical of the FUors), and Br$\gamma$ line in emission (as the EXors).  \citet{2011A&A...526L...1C} derived an effective temperature T$_{eff}$ = 3200 K, mass M$_{\star}$= 0.25 \MSun and an accretion rate of \.{M}= 1.2 $\times$ $10^{-6}$ \MSun yr$^{-1}$. This makes it one of the lowest mass YSOs with a strong outburst discovered so far. The outburst of this object began in between 2005 April and 2007 March \citep{2012ApJ...756...99F}.

\citet{2012ApJ...756...99F} found that the disc accretion rate
of V2775 Ori {  during the ourburst} increased from $\sim$ 2 $\times$ 10$^{-6}$ \MSun yr$^{-1}$
to $\sim$ 10$^{-5}$ \MSun yr$^{-1}$, about an order of magnitude
less than the canonical value for the FUors \citep[10$^{-4}$ \MSun yr$^{-1}$;][]{1996ARA&A..34..207H}.  V2775 Ori is the least luminous documented FUor outburster with a protostellar envelope (L = 28 \LSun). They modeled the source and obtained a mass for the star of M$_{\star}$= 0.5 \MSun, which is consistent with the mass obtained in \citet{2014AJ....147..140G}. {  However, they argue that the value of the mass is more likely around 0.24 \MSun, more consistent with the value given by \citet{2011A&A...526L...1C}.}  \citet{2014AJ....147..140G} estimated that the temperature of the star before and after the outburst (5600 K and
6800 K, respectively) is significantly higher than the temperatures calculated by \citet{2011A&A...526L...1C} and \citet{2012ApJ...756...99F}, about 3200-4000 K. They obtained an 
increase in the disc mass-accretion rate of one order
of magnitude larger than what \citet{2012ApJ...756...99F} derived. \citet{2014AJ....147..140G} concluded that before the outburst V2775 Ori was a Class I object, and during the outburst it is in the late stages of this class.

In this paper we present the first ALMA observations of the FUor object V2775 Ori. {  This object was observed as part of a survey of 8 FUor and EXor targets in Orion (PI: Cieza, project code: 2013.1.00710.S). The aim of this project is to investigate the origin of episodic accretion and to compare and contrast the physical characteristics of FUor/EXors with those of classical T Tauri discs. Understanding the origin of disc episodic accretion, as well as its consequences over the disc structure, is crucial for both star and planet formation \citep[see][]{2016Natur.535..258C,2017MNRAS.468.3266R, 2017MNRAS.466.3519R,2018MNRAS.474.4347C,2018MNRAS.473..879P}. The observations and data reduction are presented in Sec.~\ref{sec:obs}, followed by the results in Sec.~\ref{sec:res}, and the discussion in Sec.~\ref{sec:disc}. A summary is given in Sec.~\ref{sec:conc}.

\section{Observations and data reduction}
\label{sec:obs}

\subsection{ALMA data}

V2775 Ori has been observed three times with ALMA during Cycle-2 in band 6, as part of {  the Orion FUor/EXor survey}. Two of three observations took place on December 12$^{th}$, 2014 and April 5$^{th}$, 2015. {  The precipitable water vapor (PWV) was 0.7 and 1.3mm} for the December and April observations, respectively. The configuration of ALMA for both observing runs  was with 45 antennas (12-m of diameter) with baselines ranging from 14.6 to 348.5m. {  In this array configuration, the maximum recoverable scale of the emission is 11.4\as, which means that smooth emission at larger angular scales (e.g., from the molecular cloud or the envelope) could be missed by our observations. } ALMA {  observations} provided two spectral windows for the continuum, centered at 232.6 GHz and 218.0 GHz, 2 GHz wide, plus three narrow (59 MHz wide) bands: one centered on the \co~ (2-1, 230.5380 GHz) line (hereafter \co~), the second centered on the $^{13}$CO (2-1, 220.3987 GHz) line, and finally one centered on the C$^{18}$O (2-1, 219.5603 GHz) line. The resolution of these datasets is $\sim$ 0.8\as. The correlator setup was identical for all the observations. Ganymede and J0423-013 were used as flux calibrators, while the quasars J0538-4405 and J0607-0834 where observed for bandpass calibration. Observations of nearby phase calibrators (J0541-0541, J0532-0307 and/or J0529- 0519) were alternated with the science target to calibrate the time dependence variations of the complex gains.

The third and final observations were performed on August 30$^{th}$, 2015 with an array of 35 antennas and longer baselines of 42-1574m. For these data the resolution was higher with respect to the data from December and April, reaching $\sim$0.25\as. The PWV during the observations was $\sim$ 1mm. 

The data from all three epochs were processed and combined into a single data set. The visibility data was reduced using the Common Astronomical Software Application (CASA v4.4) package \citep{2007ASPC..376..127M}. First we concatenated the lower and higher resolutions data.  The images of the continuum and the gas line datacubes were created with the {  pipeline using the calibrated visibilities and the CLEAN } routine of CASA. Continuum subtraction was performed in the visibility domain. {  The characteristics of the beam used in the final combined dataset and analysis are detailed in Table~\ref{t:beam}. }The rms in the final continuum and line data are 0.25 mJy beam$^{-1}$ and 15 mJy beam$^{-1}$ km s$^{-1}$, respectively.

\begin{table}
\caption{{  Beam infomation of the final combined dataset.}} 
\label{t:beam}
\centering
\begin{tabular}{lllll}
\hline
\hline
   & Continuum & \co  & $^{13}$CO & C$^{18}$O \\
\hline
Major axis (\as) & 0.25 & 0.35 & 0.37 & 0.37  \\
Minor axis (\as) & 0.17 & 0.27 & 0.28 & 0.29  \\
Pos. angle (deg) & -89.09 & -89.98 & 86.70 & 87.07  \\
\hline
\end{tabular}
\end{table}

\section{Results}
\label{sec:res}

\subsection{Continuum emission}
\label{sec:dust}

The image of the continuum of V2775 Ori is shown in Fig.~\ref{f:cont}, with an integrated flux is {  $106 \pm 11$ mJy} and a peak flux is $69 \pm 0.43$ mJy beam$^{-1}$ (signal to noise ratio, SNR = 160). As the continuum image does not present any asymmetric structure, we performed a gaussian 2-dimensional fit {  with the CASA routine IMFIT and find that} the dimensions of the disc, deconvolved from the beam, are $151.0 \pm  3.6$ mas ($63 \pm 2 $ au) for the major axis and $146.5 \pm 3.7$ mas ($61 \pm 2$ au) for the minor axis. The position angle of the disc is $48 \pm 35$ deg. Assuming that the disc is circular, from the aspect ratio we calculated that the disc is almost face-on with respect to the plane of the sky, with an inclination of {  $14.0^{+7.8}_{-14.5}$} deg. The center of the disc is found in the sky location $\alpha_{2000}$ = $05^{h}42^{m}48^{s}.49$ and $\delta_{2000}$ = $-8^{\circ} 16^{\prime} 34\as.72$, slightly shifted from the ALMA phase center. {  The infrared coordinates \citep{2003yCat.2246....0C} of the star perfectly match the ALMA coordinates apart from a small shift of 0.02\as in the declination.}  A more detailed analysis and modelling of the continuum can be found in \citet{2018MNRAS.474.4347C}. 




\begin{figure}
\begin{center}
\includegraphics[width=0.5\textwidth]{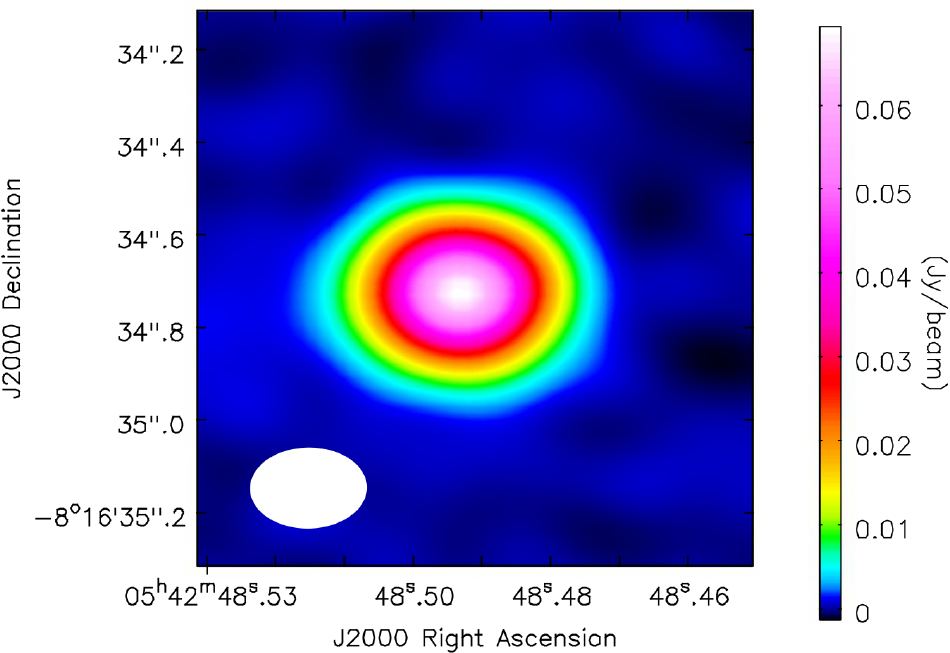}
\caption{Image of the continuum emission of the FUor V2775 Ori. The beam size is $\sim$0.25\as, shown on the lower-left corner of the image.}
\label{f:cont}
\end{center}
\end{figure}

\subsection{$^{12}$CO moments}
\label{sec:12co}
The \co~ line traces the outflow of the FUor star {  in the highest velocities, while the emission close to the system velocity may come from gas of the envelope or the disc around the central star}.  We assume that the system velocity corresponds to the velocity of the C$^{18}$O, which has faint emission in correspondence of the disk continuum emission. The velocity of the system is $3.1 \pm 0.3$ km s$^{-1}$. This value has been obtained fitting a gaussian to the peak emission of the line.

We calculated the mean intensity of the molecular lines in a circular region centered on the continuum emission, and with a diameter of 19\as ($\sim$7800 au), to include the {  whole} emission from the gas, {  after a first determination of its extension with a map integrated along the velocity channels}. The channel width used for this analysis is 0.25 km s$^{-1}$ and {  the rms is 14.46 mJy beam$^{-1}$ km s$^{-1}$}. We find three peaks in the \co~line mean flux (inside the 19\as region) vs velocity profile, as shown in Fig.~\ref{f:3_lines_plot}.   We calculated the velocities-integrated intensity (moment 0) of the gas line for three different parts, corresponding to the three peaks: the ``low-velocity'' side, the ``blue-shifted'' and the ``red-shifted'' sides, where we included only the channels exhibiting line emission. The maxima are located at a velocity v$_{LSR}$ = 0.25 km s$^{-1}$ with a mean intensity of 9.32 mJy beam$^{-1}$ for the blue part, v$_{LSR}$ = 4.78 km s$^{-1}$ with a mean intensity of 6.89 mJy beam$^{-1}$ for the low-velocity, and v$_{LSR}$ = 7.50 km s$^{-1}$ with mean intensity of 11.65 mJy beam$^{-1}$ for the red-shifted side. In Table~\ref{t:lines} we present a summary of the lines profile values.

\begin{figure*}
\begin{center}
\includegraphics[width=0.9\textwidth]{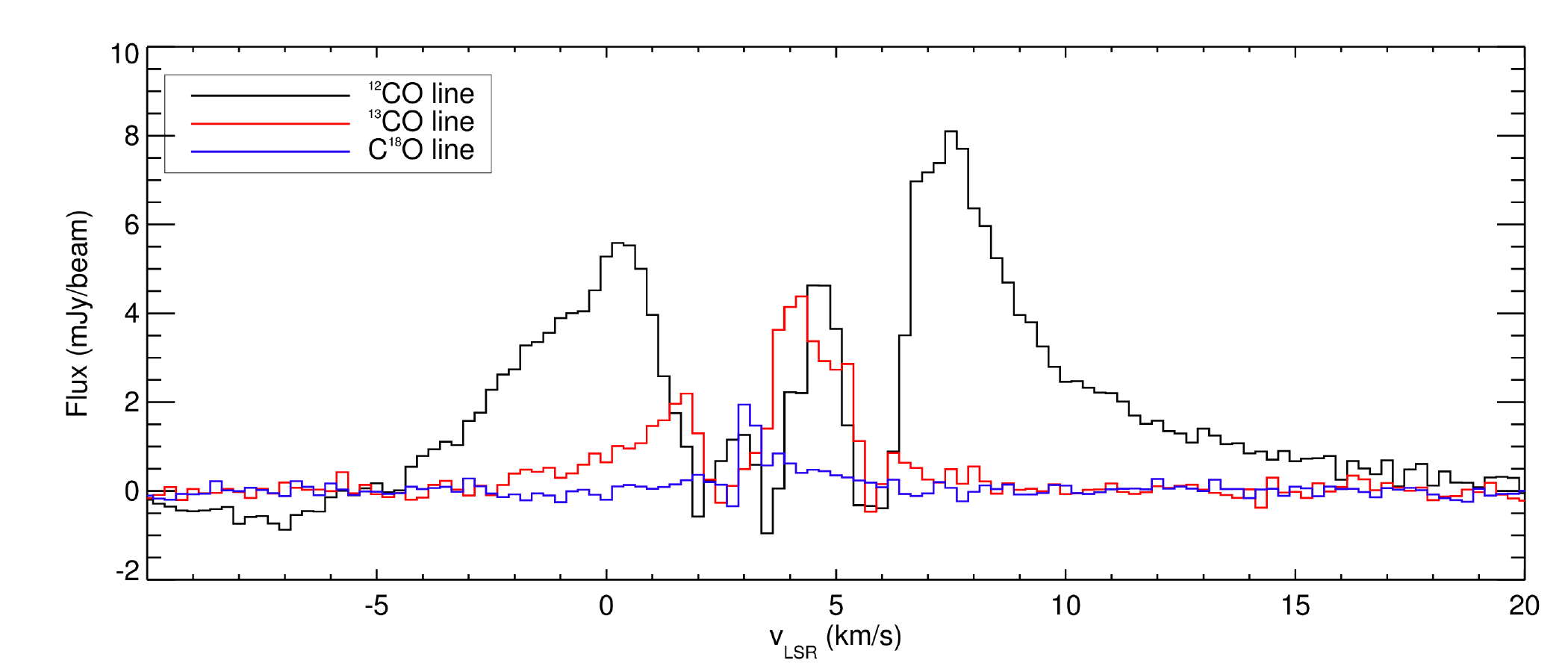}
\caption{Plot showing the mean flux (inside the 19\as circular region around the continuum) of the \co,  $^{13}$CO, and C$^{18}$O lines together versus the optical velocities. Three different peaks have been identified for the \co~line: at v$_{LSR}$ = 0.25, 4.78, and 7.50 km s$^{-1}$. Two different peaks have been identified for the $^{13}$CO line: at v$_{LSR}$ = 1.75 and 4.25 km s$^{-1}$. The intensity of the C$^{18}$O line is significantly fainter than the other two lines, its maximum is found at v$_{LSR}$ = 3.0 km s$^{-1}$.}
\label{f:3_lines_plot}
\end{center}
\end{figure*}

\begin{table}
\caption{{  Added an entry to the table.} Gas lines profiles. As the line profiles are asymmetric we refer to the peak of the emission. The velocity v$_{LSR}$ and the mean flux at the peak of each side is indicated as v$_s$ and $F_s$, respectively.   } 
\label{t:lines}
\centering
\begin{tabular}{cccccc}
\hline
\hline
 &  Side & Range & v$_s$  & $F_s$ & v$_s$ - v$_{\mathrm{C^{18}O}}$  \\
  &    &(km s$^{-1}$) &  (km s$^{-1}$)  & (mJy beam$^{-1}$) & (km s$^{-1}$)\\
\hline
\parbox[t]{2mm}{\multirow{3}{*}{\rotatebox[origin=c]{90}{\co}}} & B &-5 $\sim$ 2  &  0.25 & 9.32  & { -2.85}\\
   &   L  & 4 $\sim$ 5.25   & 4.78  & 6.89 & { 1.68} \\
& R & 6 $\sim$ 16 & 7.5  & 11.65  & { 4.4}\\
\hline
\parbox[t]{2mm}{\multirow{3}{*}{\rotatebox[origin=c]{90}{$^{13}$CO}}}  & B  &  -2.5 $\sim$ 2.5& 1.75 & 3.36 & { -1.35} \\
& R  &  2.75 $\sim$ 7.5 & 4.25 & 7.30 & { 1.15} \\
 &&&&& \\
\hline
\parbox[t]{2mm}{\multirow{3}{*}{\rotatebox[origin=c]{90}{C$^{18}$O}}} &&&& \\
 & L   &  2.75 $\sim$ 4 & { 3.1} & 1.94 & 0 \\
&&&&& \\
\hline
\end{tabular}
\end{table}

 The moments 0 maps of \co~ associated with the above velocity ranges are shown in Fig.~\ref{f:12co_mom0}, and a more detailed velocity channel map is shown in {  the online material}.

\begin{figure*}
\begin{center}
\includegraphics[width=\textwidth]{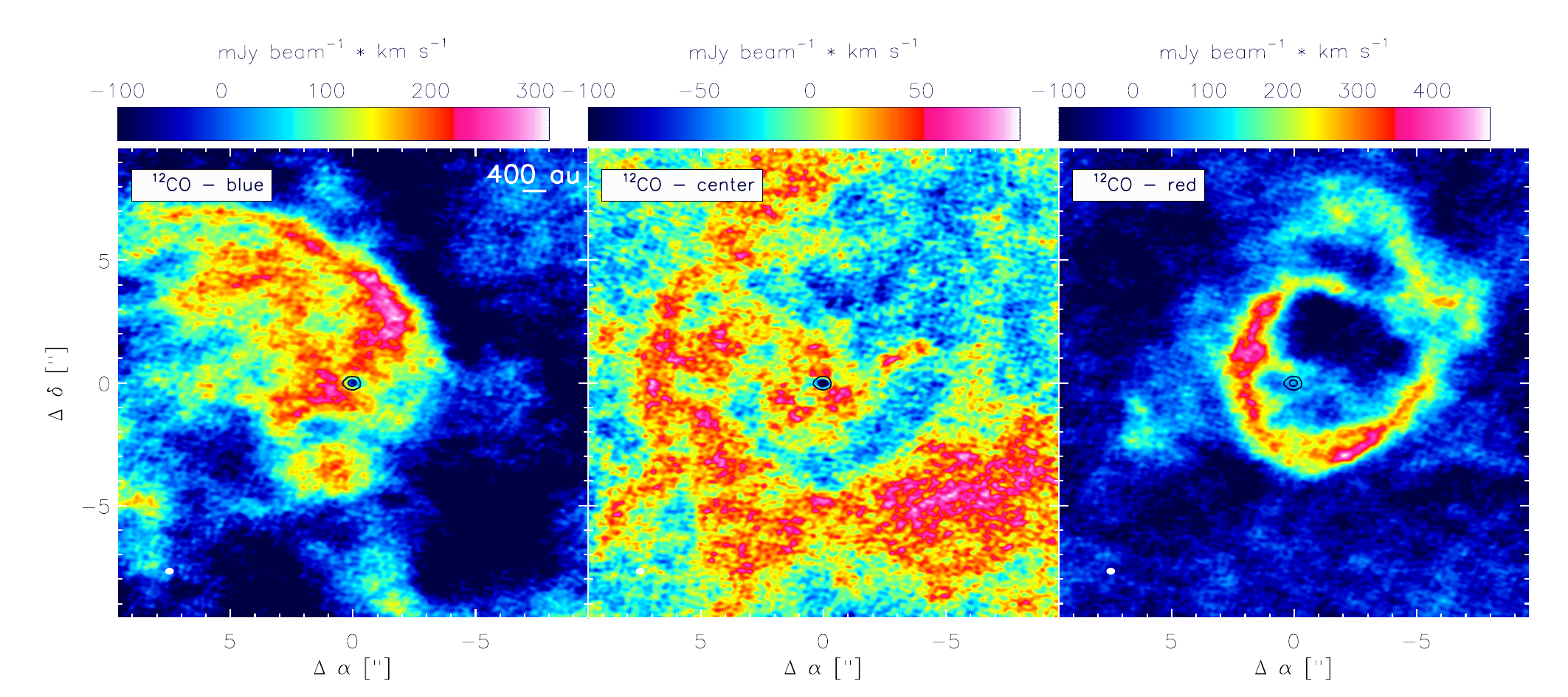}
\caption{Moments 0 maps of the \co ~line. The blue-shifted side of the velocities (left), the low-velocities side (center), and the red-shifted side (right) are shown. {  We divided the three parts following the velocity ranges listed in Table~\ref{t:lines}.} The beam size is $\sim$0.25\as, shown on the lower-left corner of the image. {  The continuum is shown in blue contour map.} }
\label{f:12co_mom0}
\end{center}
\end{figure*}

The blue-shifted side indicates that the gas approaching along the line-of-sight is concentrated in a massive shell that extends from the center-west to the north-east of the image. The extended cavity goes from a distance of $\sim$ 1344 au (3.2\as) from the center of the disc to $\sim$ 4200 au (10\as). The rms of the blue-shifted moment 0 image is 60 mJy beam$^{-1}$ km s$^{-1}$.



The low-velocities channels show a diffuse and low flux emission distributed in the shape of a half-ring, centered in the bottom-left side of the image. The radius of the half-ring is $\sim$ 3000 au. The red-shifted side shows a compact ring of strong emission, slightly decentered ($\sim$ 600 au) from the disc position, with a radius of $\sim$ 1500 au.

\subsection{$^{13}$CO moments}
\label{sec:13co}
The $^{13}$CO has a lower optical depth with respect to the \co~ and it {  is a tracer of medium density material \citep[$\sim$ 10$^{4}$ cm$^{-3}$;][]{2005ism..book.....L}}. {  Similar to that of the \co~emission}, the intensity of the $^{13}$CO versus the velocity presents more than one peak. The plot of the {  line intensity versus velocity is displayed} in Fig.~\ref{f:3_lines_plot}. Similar to the case of \co~data, we divided the datacube in two parts: the red-shifted side and the blue-shifted side. The maxima are located at v$_{LSR}$ = 1.75 km s$^{-1}$ with a mean intensity of 3.36 mJy beam$^{-1}$ for the blue part, and v$_{LSR}$ = 4.25 km s$^{-1}$ with a mean intensity of 7.30 mJy beam$^{-1}$ for the red-shifted part.  {  The rms is 18.07 mJy beam$^{-1}$ km s$^{-1}$}. The moments 0 map for the two sides are shown in Fig.~\ref{f:13co_mom0} {  and a more detailed velocity channels map is shown in the online material}.

\begin{figure*}
\begin{center}
\includegraphics[width=\textwidth]{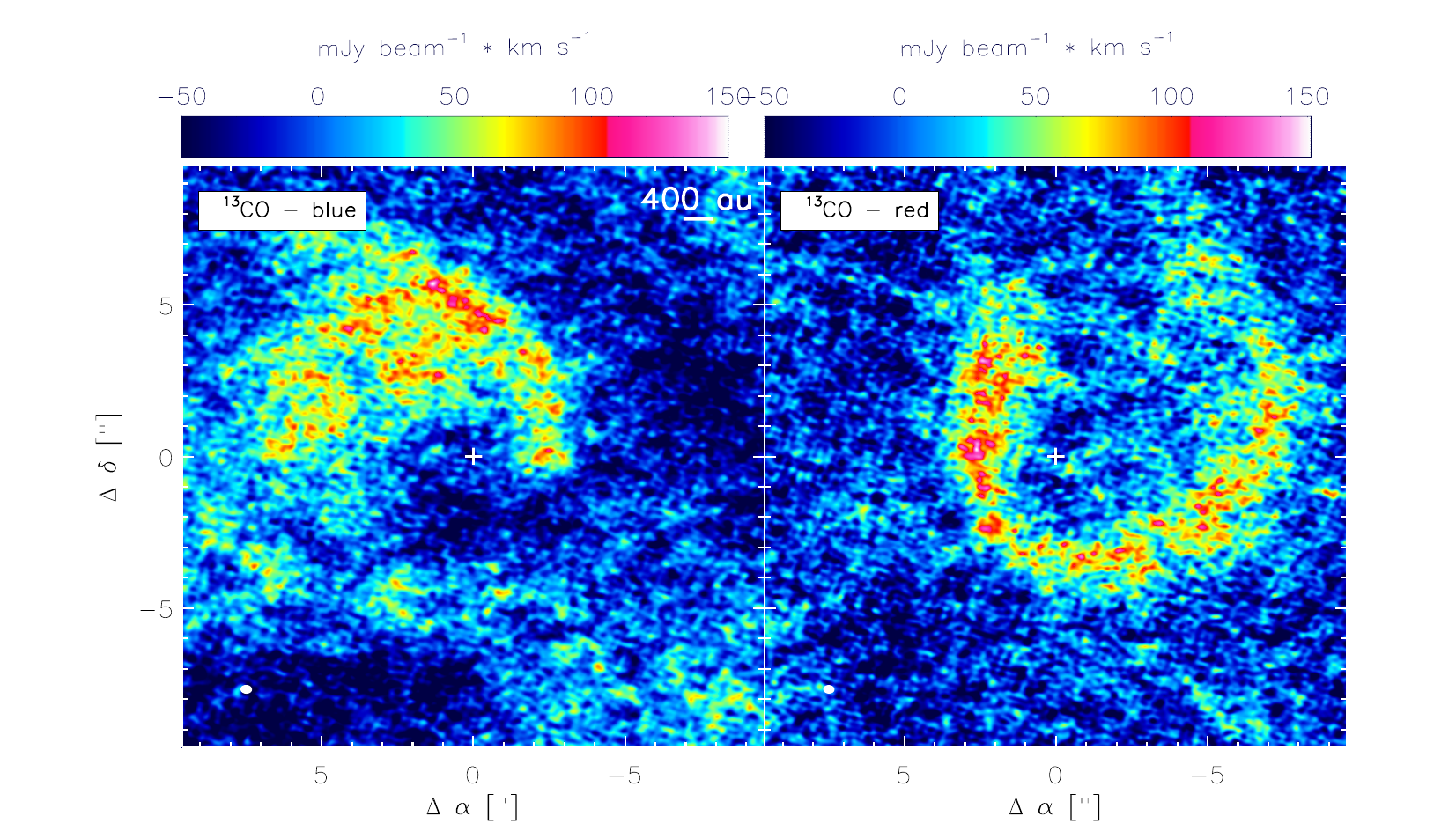}
\caption{Moments 0 maps of the $^{13}$CO line. The blue-shifted side (left) and the red-shifted part (right) are shown. {  The velocity ranges are listed in Table~\ref{t:lines} and the beam size is $\sim$0.25\as which is shown in the lower left corner of the images.  The position of the continuum emission is indicated as a white cross.}}
\label{f:13co_mom0}
\end{center}
\end{figure*}

The blue-shifted part {  of the $^{13}$CO line is similar to that of the blue side of the \co~ emission}. The gas is located in the same extended shell, as shown in Fig.~\ref{f:co12_co13} (left), where we overplotted with contour lines the emission of the $^{13}$CO on the blue-shifted moment 0 map of the \co~line. The red-shifted part of the gas spatially coincides with the ring seen in the red-shifted part of the \co ~line, as shown in Fig.~\ref{f:co12_co13} (right), but with velocities comparable with the low-velocities side of the \co.  This unusual behavior has also been noticed in the outflow of the object HH 46/47 \citep{2016ApJ...832..158Z}. In fact, as we can see from the plot in Fig.~\ref{f:3_lines_plot}, the two peaks of the $^{13}$CO emission are a scaled (in flux) version of the profile of the \co~emission, {  with closer velocities to the systemic velocity}. The distribution along the velocity axis is the same for both emission lines, and the correspondent low-velocity side is apparently too weak to be detected.

{ When well mixed, \co~ and $^{13}$CO trace the same material. However, that does not mean that the emerging emission lines appear the same. This is due to an optical depth effect rather than a different density field. For example in an outflow, at higher velocity, the column density of the material is not high, and both the isotopologues are optically thin. In such a case, the $^{13}$CO emission may be too faint to be detected, and we only see the outflow structure in the \co~ emission. At low velocities, the outflow material has much higher column density and it starts to mix up with the cloud material. For optically thin tracers like $^{13}$CO, outflow structure can still be distinguished from the cloud emission because the outflow material piled up along the outflow cavity with higher column density. In the case of interferometric observation, the cloud emission may be completely filtered out and only the outflow cavities are seen. Meanwhile, \co~ quickly becomes optically thick at these velocities, therefore it is hard to distinguish outflow emission from the cloud emission. In the case of our observations the maximum angular scale is 11\farcs4 and the cloud emission may be filtered out. }

The \co~shows absorption in some places where the $^{13}$CO peaks, suggesting multiple cones/rings along the line of sight with an ordered velocity gradient.

\begin{figure*}
\begin{center}
  \includegraphics[width=\textwidth]{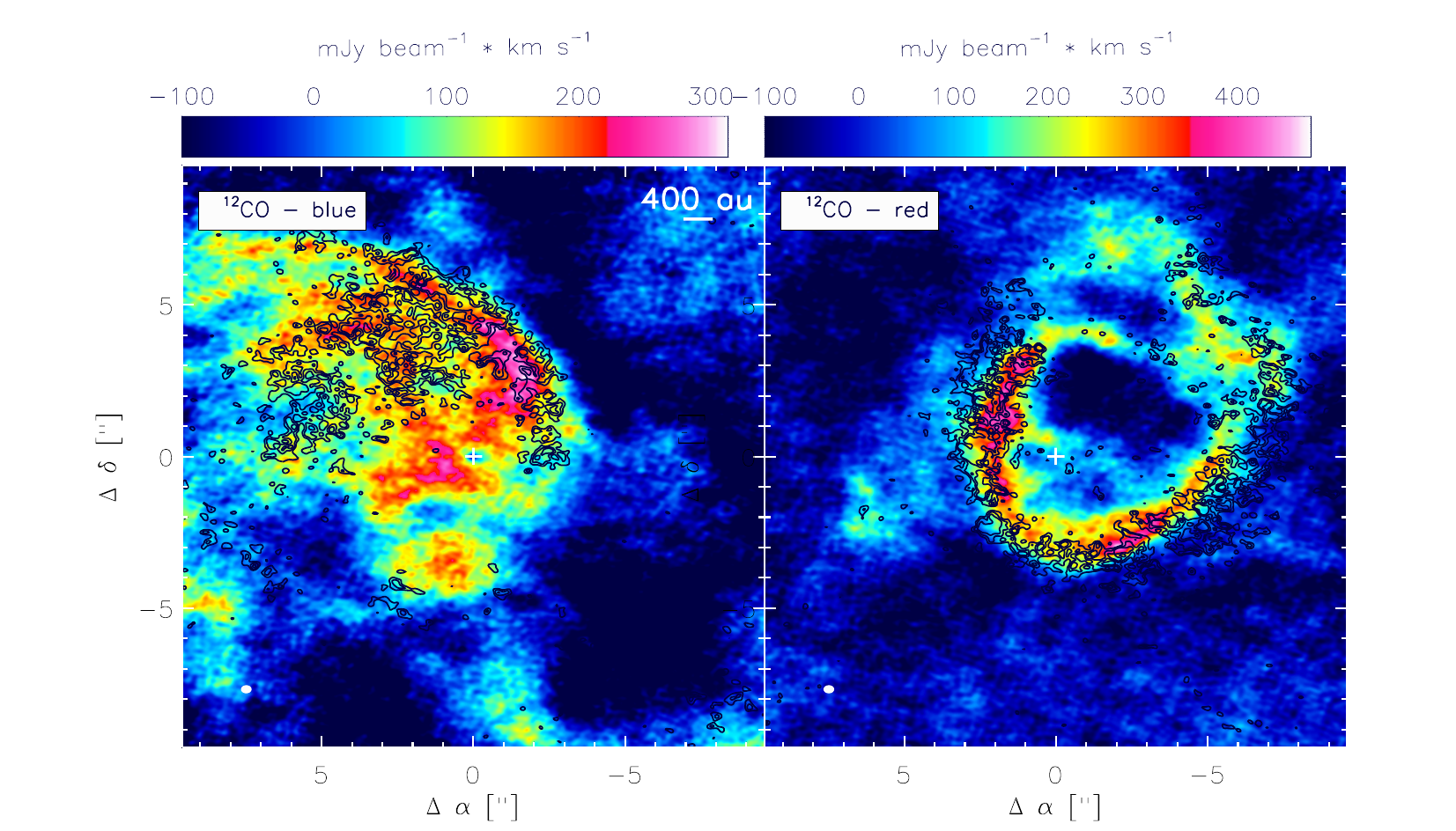}
\caption{{\it Left}: Image showing the blue-shifted side of the $^{12}$CO line with the black contour map of the blue-shifted side of the $^{13}$CO line. Both tracers are located in the same shell. The continuum is shown {  as a white cross}. {\it Right}: Image showing the red-shifted side of the $^{12}$CO line with the blackcontour map of the red-shifted side of the $^{13}$CO line. Both tracers are located in the ring. The star is located on the position of the white cross. The beam size is $\sim$0.25\as, shown on the lower-left corner of the image. }
\label{f:co12_co13}
\end{center}
\end{figure*}



\subsection{C$^{18}$O moments}
\label{sec:18co}

The C$^{18}$O emission is found in one peak of the intensity versus velocities plot, corresponding to the low-velocities of the \co. {  The rms is 10.79 mJy beam$^{-1}$ km s$^{-1}$}. The moment 0 map for the C$^{18}$O line is shown in Fig.~\ref{f:18co_mom0}. The emission of this line is significantly fainter with respect to the lines of the \co~ and $^{13}$CO, as we show in Fig.~\ref{f:3_lines_plot}. The emission peaks at the location of the disc, as indicated via the continuum emission, and the spectrum is narrow, consistent with the low inclination of $\sim$ 14 deg from the continuum fitting.

\begin{figure*}
\begin{center}
  \includegraphics[width=0.45\textwidth]{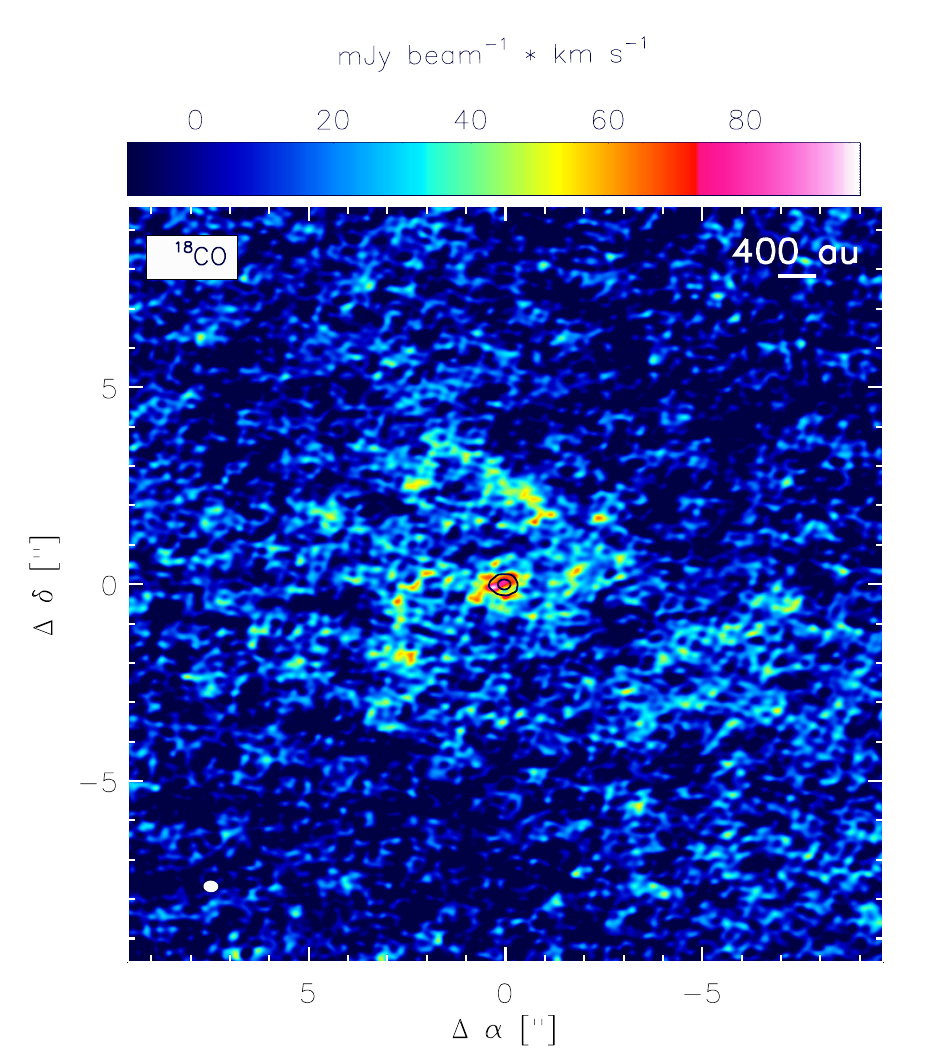}
  \includegraphics[width=0.45\textwidth]{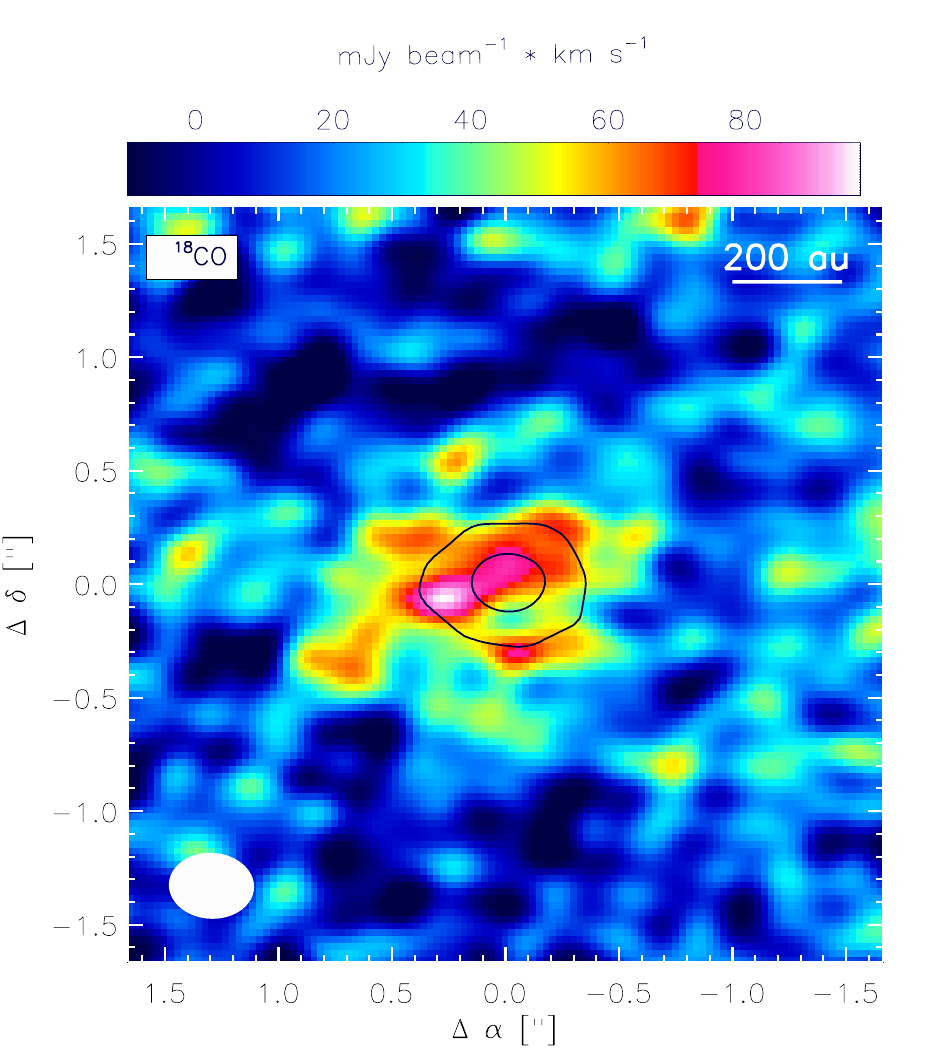}
\caption{Moment 0 map of the C$^{18}$O line. On the right side a zoom around the center of the disc is shown. The beam size is $\sim$0.25\as, shown on the lower-left corner of the image. {  The continuum is shown as blue contours.} }
\label{f:18co_mom0}
\end{center}
\end{figure*}

\subsection{Disc gas mass estimate}

The strong \co~ and $^{13}$CO emission indicates extended structure related to
the protostellar outburst and its effect on the surrounding envelope.
Fig.~\ref{f:3_lines_plot} shows strong absorption from this envelope and this
precludes a detection of these lines in the disc.
The C$^{18}$O line also shows some extended emission but the
absorption is much weaker due to its lower optical depth and we
find a compact source centered on the continuum source. {  The flux of this compact source is 0.15 Jy km s$^{-1}$. The value of the dust temperature is set as 20 K. The gas temperature is a parametric function of radius and vertical height above the midplane. We use this to estimate a lower limit to the disc gas mass of 3 \MJup based on the models of \citet{2014ApJ...788...59W}.}

The gas mass may be considerably higher if the disc is so cold that most of the gas is below 20 K and the CO is frozen out.  The dust mass is estimated to be $\sim 1.6\times 10^{-3}\,M_\odot$
for ``standard'' dust opacity, $\kappa$ = $2.2\,{\rm cm}^2\,{\rm g}^{-1}$,
and temperature, $T_{\rm dust}$ = $20$\,K \citep[e.g.][]{2011ARA&A..49...67W}.
Therefore, even if the gas mass is underestimated by an order
of magnitude, the gas-to-dust ratio appears to be much lower
than the ISM ratio of 100.
\citet{2014ApJ...788...59W} and \citet{2016ApJ...828...46A} also found
very low CO isotopologue line-to-continuum ratios in Taurus and
Lupus discs. If we assume the typical gas-to-dust ratio of $\sim$20 measured in Taurus and Lupus discs, based on [$^{13}$CO/C$^{18}$O] compared to the continuum, the gas mass would be $\sim$30 \MJup. Our finding here suggests the loss of gas or chemical effect that creates such a low C$^{18}$O line to continuum ratio is very rapid and perhaps may be related to the strong outbursts of early protostellar evolution.

{  Indeed, comparing our results with the archetype of the EXor class, EX Lup, \citet{2015ApJ...798L..16B} found that after the strong outburst of 2008 the CO decreased dramatically. This can be caused by a rapid depletion of gas accumulated beyond the disc corotation radius during quiescent periods. \citet{2016ApJ...821L...4K} found that EX Lup has a modest CO depletion. Using the $^{13}$CO line, and the canonical $10^{-4}$ CO-to-H$_{2}$ abundance ratio, they estimate a total disc mass of $2.3\times 10^{-4}$ \MSun. This results can be a factor 10-100 lower than the actual value taking into account the CO depletion. }

\subsection{Outflow gas parameters}
\begin{table*}
\caption{Column density, mass, momentum and kinetic energy of the outflow.} 
\label{t:mass}
\centering
\begin{tabular}{cccccc}
\hline
\hline
Molecule &  Side & Column density ($10^{20}$ m$^{-2}$) & Mass ($10^{-3}$\MSun)  & Momentum ($10^{-3}$\MSun km s$^{-1}$)  & Energy ($10^{40}$ erg)  \\
\hline
\parbox[t]{2mm}{\multirow{3}{*}{\rotatebox[origin=c]{90}{\co}}} & Blue-shifted  & {  2.8} & {  1.8}  & {  6.9} & {  29} \\
& Red-shifted & {  6.2}  & {  10.5}  & {  53} & {  280} \\
& Total & {  9.0} & {  12.3}  & {  60.0} & {  310} \\
\hline
\parbox[t]{2mm}{\multirow{3}{*}{\rotatebox[origin=c]{90}{$^{13}$CO}}}  & Blue-shifted  &  0.7 & 0.2 & 0.3 & 0.8 \\
& Red-shifted  & 1.3 & 1.0 & 1.3& 2.0 \\
& Total & 2.0 &  1.1 & 1.6 & 2.8  \\
\hline
\end{tabular}
\end{table*}

{  Column density, mass, momentum, and kinetic energy of the expanding gas are listed in Table~\ref{t:mass}}. To calculate these quantities we used the formalism presented in \citet{2015ApJ...798...85P}, {  and applied it to both the \co~ and $^{13}$CO isotopologues.  We did not determine these properties for the C$^{18}$O gas because its emission was too weak.} The correction for the optical depth \citep[see, e.g,][]{2014ApJ...783...29D} cannot be applied for this dataset, as only a few velocity channels present emission from both the isotopologues. The values presented are then lower limits of these quantities. We adopt in this paper an abundance of the \co~relative to H$_2$ of 10$^{-4}$, and a relative abundance between \co~and $^{13}$CO of 62 \citep{1993ApJ...408..539L}. The excitation temperature, assumed constant along the line of sight, has a value of T$_{ex}$ = 50 K in this analysis. The beam filling factor is assumed to be 1. The total mass of the gas is defined as $M$ = $\sum_{v}${M(x,y,v)}, the momentum surface density is defined as $P$ = $\sum_{v}$ M(x,y,v)v, and the kinetic energy surface density as $E$ = $\sum_{v}${ M(x,y,v)v$^2$/2}, where v = v$_{LSR}$-v$_{\mathrm{C^{18}O}}$, and M(x,y,v) is the mass surface density calculated for each velocity channel.

The values found for our object have the same order of magnitude of other protostars presented in literature: for example, \citet{2006ApJ...646.1070A} presented observations of nine low-mass, Class 0, I, and II sources with detected outflows. They found outflows' masses in the range of 6--150 $\times 10^{-3}$ \MSun, with momenta in the range 4--119 $\times 10^{-3}$ \MSun km s$^{-1}$, and kinetic energy of 2--484 $\times 10^{40}$ ergs. These values are very similar to those reported here for V2775 Ori, keeping in mind that this source is seen face-on, so only a small portion of the outflow is detected. Similar values are found for HL Tau system \citep{2016MNRAS.460..627K}.

Due to the inclination of the outflow, we cannot estimate the size of the outflow along the Z axis (towards us), and consequently the age of the system. We can speculate that the component of the velocity of the gas along the plane of the sky should be lower than the component along the Z axis, as \citet{2006ApJ...646.1070A} suggests, young Class 0 object have highly collimated outflows. If this is the case, taking the highest velocity of the gas, i.e. 4.5 km s$^{-1}$ for the red-shifted part and assuming it constant, and the radius of the ring of 1500 au, the kinetic age of the outflow would be less than $\sim$ 1600 yr. Another independent way to estimate the age of the outburst is to assume that the axis of the hourglass is normal to the plane of the disc, which is inclined by $\sim$14 deg. Taking into account that the center of the red-shifted ring is shifted by 600 au, the hourglass half axis would be of 2500 au. In this case the opening angle of the outflow is of 30 deg, and the age is about 2600 yr. As the disc is almost face-on, the uncertainties on these values are high. Nevertheless, the same order of magnitude is found for HL Tau \citep[2600 yr;][]{2016MNRAS.460..627K} and HBC 494 \citep[2900 yr;][]{2017MNRAS.466.3519R}. The outflow of V2775 Ori is much more collimated than the one of HBC 494 \citep[$\sim$ 150 deg][]{2017MNRAS.466.3519R}. \citet{2006ApJ...646.1070A} suggest that there is a correlation between the opening angle of the outflow and the age of the system: the opening angle is increasing for more evolved system. The age of 0.1 Myr for V2775 Ori is compatible with the value of the opening angle of 30 deg.


\begin{figure}
\begin{center}
\includegraphics[width=0.45\textwidth]{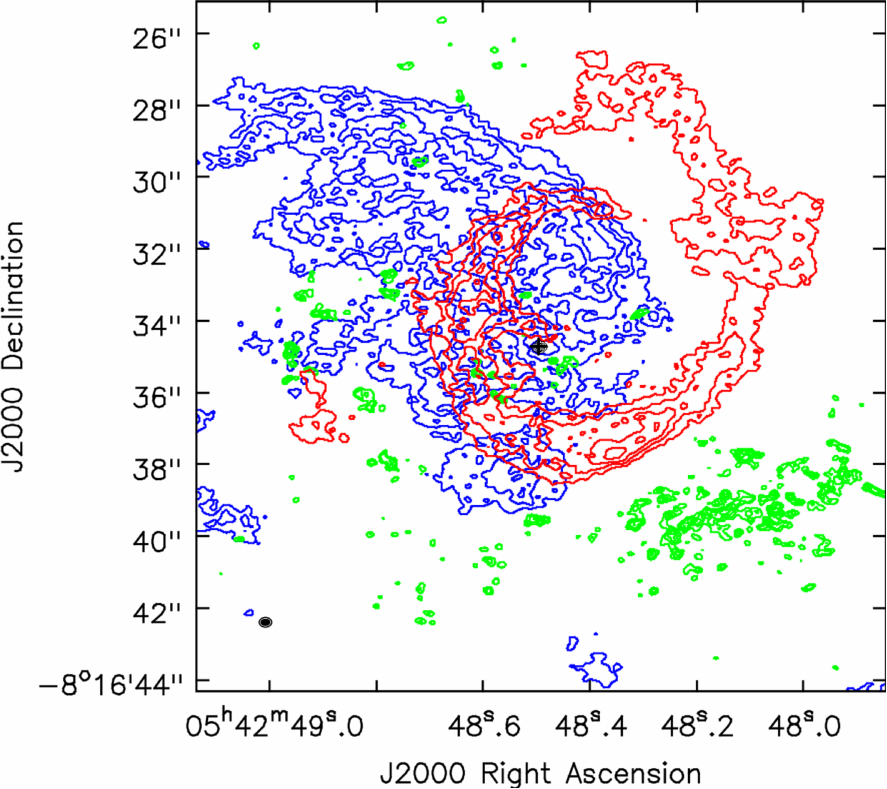}
\caption{Three colours {  to simulate a 3D image} showing the blue-shifted, central, and red-shifted part of the \co~line emission. The three different zones are shown in blue, green, and red contours for the blue-shifted, low-velocities, and red-shifted parts respectively. The contours represent 0.2, 0.4, 0.6, 0.8 step levels of 0.04 Jy beam$^{-1}$. The continuum emission is located where the black cross is. The beam size is $\sim$0.25\as, shown on the lower-left corner of the image. {  A cartoon representation can be found in Fig.~\ref{f:cartoon}}.}
\label{f:12co_3d}
\end{center}
\end{figure}

\begin{figure}
\begin{center}
\includegraphics[width=0.45\textwidth]{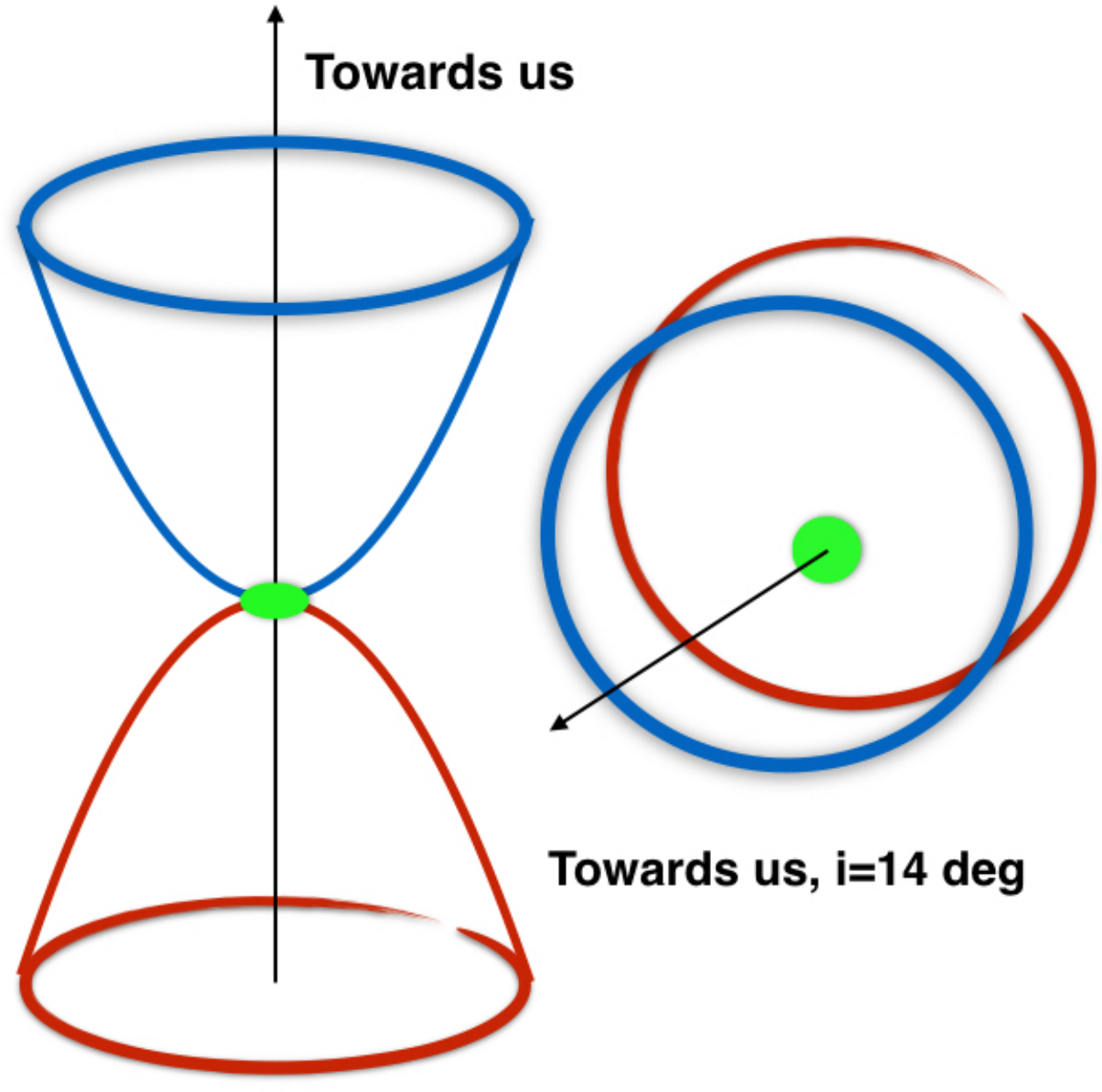}
\caption{Cartoon sketch of the system V2775 Ori. The palette of colours is the same of Fig.~\ref{f:12co_3d} to indicate different regions of the outflow. The disc is not to scale. The arrows indicate the axis of the hourglass. }
\label{f:cartoon}
\end{center}
\end{figure}

\section{Discussion}
\label{sec:disc}
The \co~ emission traces cool material and it is optically thick. The emission measured of this line presents three peaks for this FUor object: one in the blue-shifted side of the velocities, one in the low-velocities side, and one in the red-shifted velocities. This indicates that the gas is expanding from the disc, and as it is seen face-on, we expect to look through the outflow of this object. Indeed, in the blue and red sides of the velocities, the gas is located in circular shapes and aligned with respect to the disc (Fig.~\ref{f:12co_3d}). That means that the axis of the ``hourglass'' shape is almost aligned with the line-of-sight. A cartoon sketch of the system is presented in Fig.~\ref{f:cartoon}. On the other hand, we do not see the walls of this hourglass. {  The emission is concentrated in the shell (approaching us) and the ring shape (moving away from us) but there is almost no connection between these regions and the disc.} The last outburst for this object took place between 2005 and 2007. If we assume a constant velocity for the gas of 5 km s$^{-1}$ \citep{2016MNRAS.460..627K}, the ejected gas should be located at $\sim$12 au away from the star by now. The outburst seen in our dataset cannot be that close to the star, which suggests that the \co~is a remnant left from a previous outflow $\sim$2000 years old, given the dynamical age estimated in the previous section. The shells apparent in the data represent cavities in the molecular cloud excavated by the outflow of the FUor. This scenario is similar to the one presented by \citet{2005ApJ...632..941Q}, where they speculate that the cavities seen in the molecular cloud NGC 1333 are the fossils of ancient outflows of pre-main sequence stars.

Since accretion and outflow rates are intimately connected, an episodic outflow scenario is perfectly consistent with the episodic accretion paradigm \citep{2014prpl.conf..387A}. In fact, clear signatures of such episodic molecular outflows have recently been imaged by ALMA \citep{2015Natur.527...70P}. 

In our object the blue-shifted emission of the $^{13}$CO line is located at the same position of the blue-shifted part of the \co, same for the red-shifted part but with velocities comparable to the low-velocities side of the \co. We can interpret this phenomenon as emission from the \co~ and $^{13}$CO lines tracing the low- and medium-density gas, respectively. Then, the $^{13}$CO is moving with lower velocities because of the higher density of the gas that it is tracing. Again, in this emission line we do not see the walls of the cone-shape outflow, but only two symmetrical circular features, expanding away from the disc.

The C$^{18}$O is optically thin, and thus probes deeper regions. \citet{2015Natur.527...70P} suggest that the C$^{18}$O may trace the envelope of the circumstellar disc. {  We detected very faint gas emission from the disc of V2775 Ori and we use this to estimate the velocity of the system.}  If we assume that the velocity of the C$^{18}$O is the velocity of the system we can estimate that the blue-side emission of the \co~line is approaching us with a velocity of { 2.85} km s$^{-1}$, while the red-side is expanding away from us with a velocity of { 4.4} km s$^{-1}$ (see Table~\ref{t:lines}). {  The non-symmetrical velocities may be explained by the fact that the gas is expanding into a space with material of different densities. }Similar velocities but slightly lower are found for the outflow of HBC 494 \citep{2017MNRAS.466.3519R}. The higher density gas traced by the $^{13}$CO is moving slower with symmetric velocities of { $\sim $1} km s$^{-1}$. 

\section{Summary}
\label{sec:conc}

We observed the FUor object V2775 Ori with {  ALMA and obtained a resolved image of the disc continuum emission with a resolution of 0.25\as.} The disc is seen nearly face-on, with an inclination of $\sim$14 deg. The characteristic radius of the disc is $\sim$ 30 au. We also obtained three narrow bands centered on the \co~ (2-1, 230.5 GHz) line, the $^{13}$CO (2-1, 220.4 GHz) line, and the C$^{18}$O (2-1, 219.6 GHz) line.  The \co~ and $^{13}$CO emission features show similar structures expanding from the disc: the blue-shifted side of the gas is concentrated in an extended shell, while the red-shifted side is concentrated in a compact ring. The denser gas, traced by $^{13}$CO emission, is moving with lower velocities with respect to the \co, but is tracing the same structures. { Since we cannot see the walls of this ``hourglass'', we can interpret these structures as cavities created by an ancient outflow.  We argue that this structure is a fossil of an outburst occurring before the previous outburst in 2005-2007.} Similar cavities have been already seen in other molecular clouds \citep[see, e.g.,][]{2005ApJ...632..941Q}. The previous outburst, assuming than the outflow is collimated (i.e. with an opening angle $<$ 45 deg), had occurred less than 1600 yr ago. The opening angle of the outburst should be of the order of 30 deg, given the decentered position of the red-shifted ring of the \co.



\section*{Acknowledgements}
This paper makes use of the following ALMA data:
ADS/JAO.ALMA\#2013.1.00710.S . ALMA is a partnership of
ESO (representing its member states), NSF (USA) and NINS (Japan),
together with NRC (Canada), NSC and ASIAA (Taiwan), and
KASI (Republic of Korea), in cooperation with the Republic of
Chile. The Joint ALMA Observatory is operated by ESO, AUI/NRAO
and NAOJ. The National Radio Astronomy Observatory is a facility
of the National Science Foundation operated under cooperative
agreement by Associated Universities, Inc.
A.Z., L.A.C., acknowledges support from the Millennium Science Initiative (Chilean Ministry of Economy),
through grant ``Nucleus RC130007''. L.A.C. also acknowledges support from FONDECYT-CONICYT grant \#1140109. H.C. acknowledges support from the Spanish Ministerio de Econom\'ia y Competitividad under grant AYA2014-55840P. S.P. acknowledges financial support by FONDECYT grant \#3140601. D.A.P. acknowledges FONDECYT grant \#3150550.




\bibliographystyle{mnras}
\bibliography{v2775} 






\bsp	
\label{lastpage}
\end{document}